\documentclass[conference,fleqn,times]{IEEEtran}
\IEEEoverridecommandlockouts
\usepackage{cite}
\usepackage{amsmath,amssymb,amsfonts}
\usepackage{algorithmic}
\usepackage{graphicx}
\usepackage{textcomp}
\usepackage{balance}
\usepackage{xcolor}
\usepackage{multirow}
\usepackage{booktabs}
\usepackage[normalem]{ulem}
\usepackage{hyperref}
\usepackage{algorithm}
\usepackage{algorithmic}
\usepackage{soul}
\useunder{\uline}{\ul}{}

\def\BibTeX{{\rm B\kern-.05em{\sc i\kern-.025em b}\kern-.08em
    T\kern-.1667em\lower.7ex\hbox{E}\kern-.125emX}}
\begin{document}

%\title{A scalable approach to Minimum Variance Distortionless Response beamforming for large arrays 
\title{A Scalable MVDR  Beamforming  Algorithm That is Linear in the Number of Antennas

\thanks{Sponsored by the Office of Naval Research via the NRL Base Program. CAM was supported in part by AFOSR Young Investigator Program Award \#FA9550-22-1-0208 and ONR award N000142312752. AG and RD were supported in part by ONR Award N000142312086, and by a gift from Dolby.
Distribution Statement A: Approved for public release. Distribution unlimited.}
}

\author{\IEEEauthorblockN{Sanjaya Herath\IEEEauthorrefmark{1}, Armin Gerami\IEEEauthorrefmark{1}, Kevin Wagner\IEEEauthorrefmark{2}, Ramani Duraiswami \IEEEauthorrefmark{1},  Christopher A. Metzler\IEEEauthorrefmark{1} \\
Email: sanjayah@umd.edu, agerami@umd.edu, kevin.wagner@nrl.navy.mil, ramanid@umd.edu, metzler@umd.edu}
\IEEEauthorblockA{\IEEEauthorrefmark{1} University of Maryland, College Park, MD \hspace*{0.5in} \IEEEauthorrefmark{2} Naval Research Laboratory, Washington, DC} 
}

\maketitle

\begin{abstract}

% Answer the following questions in the abstract:
% What are you trying to do? Articulate your objectives using absolutely no jargon.
% How is it done today, and what are the limits of current practice?
% What is new in your approach and why do you think it will be successful?
% Who cares? If you are successful, what difference will it make?

The Minimum Variance Distortionless Response (MVDR) beamforming technique is widely applied in array systems to mitigate interference. However, applying MVDR to large arrays is computationally challenging; its computational complexity scales cubically with the number of antenna elements.
In this paper, we introduce a scalable MVDR beamforming method tailored for massive arrays. Our approach, which is specific to scenarios where the signal of interest is below the noise floor (e.g.,~GPS), leverages the Sherman-Morrison formula,  low-rank Singular Value Decomposition (SVD) approximations, and algebraic manipulation. Using our approach, we reduce the computational complexity {\em from cubic to linear in the number of antennas}.
We evaluate the proposed method through simulations, comparing its computational efficiency and beamforming accuracy with the conventional MVDR approach. Our method significantly reduces the computational load while maintaining high beamforming accuracy for large-scale arrays.
This solution holds promise for real-time applications of MVDR beamforming in fields like radar, sonar, and wireless communications, where massive antenna arrays are proliferating.%becoming more prevalent.
\end{abstract}
\begin{IEEEkeywords}
large arrays, MVDR,  Scalable beamforming
\end{IEEEkeywords}
%\vspace*{-4pt}
\section{Introduction}
%  Importance of large-scale arrays
Large-scale arrays are widely deployed in various applications such as radar, sonar, and wireless communications \cite{i_largescale}. These arrays consist of a large number of antennas ranging from hundreds to thousands. The large number of antennas in the array provides high spatial resolution and the ability to suppress interference. The large array size is particularly crucial in applications requiring precise localization and tracking of targets in the presence of interference. However, as arrays scale up, the associated signal processing tasks, especially beamforming, become more complex and computationally demanding \cite{i_large_issues}.

%  Beamforming 
Beamforming is a signal processing technique used in array systems to direct the reception or transmission of signals towards a specific direction \cite{i_beamforming2}. By adjusting the phase and amplitude of the signals received or transmitted by each antenna, beamforming allows for focusing the array's sensitivity towards a desired direction while suppressing interference from other directions \cite{i_beamforming}. Beamforming is utilized in narrowband applications such as radar target detection, as well as in broadband applications such as sound scene analysis \cite{i_beamforming4}. 
Several beamforming techniques have been developed over the years, each with its own advantages and limitations \cite{i_beamforming3}. Among these, MVDR beamforming is a popular beamforming technique for its ability to enhance the signal-to-inteference-plus-noise ratio (SINR) \cite{i_mvdr}. MVDR achieves this by minimizing the total output power of the array while maintaining a distortionless response to the signal of interest. This makes MVDR beamforming particularly useful in applications where the signal of interest is weak, and the interference is strong, such as in cluttered or interference-rich environments.

%  Problem of MVDR beamforming in large arrays
While MVDR beamforming is effective in smaller arrays, its application to large-scale arrays introduces significant challenges. The primary challenge is the computational complexity associated with estimating the covariance matrix and inverting the covariance matrix. For large arrays, the covariance matrix grows quadratically with the number of antennas, and its inversion grows cubically \cite{rl_smi_mvdr}. As a result, real-time implementation of MVDR beamforming in large arrays becomes computationally infeasible, especially in scenarios where the covariance matrix needs to be updated frequently, such as in dynamic environments with moving targets and interferers.

%  Proposed method
In this paper, we propose a scalable approach to MVDR beamforming for large arrays, specifically when the desired signal is below the noise floor. The proposed method leverages the Sherman-Morrison formula \cite{i_sherman} and low-rank Singular Value Decomposition (SVD) approximations \cite{i_svd} to reduce the computational complexity of the MVDR beamforming. The proposed method allows for real-time implementation of MVDR beamforming in large arrays by updating the inverse of the covariance matrix in $O(MK^2)$ operations per time step, where $M$ is the number of antennas and $K$ is a number much smaller than $M$. We evaluate the performance of the proposed method in terms of computational complexity and beamforming accuracy using simulations and compare it with the conventional MVDR beamforming method.

%\vspace*{-6pt}
\section{Related Work}

%  Existing methods
Several methods have been proposed to address the computational complexity of beamforming in large arrays. They can be broadly categorized into three groups: algorithmic approaches, distributed or parallel approaches, and deep learning-based approaches.

%  Algorithmic approaches
Algorithmic approaches aim to reduce the computational complexity of MVDR beamforming by approximating the covariance matrix or using iterative algorithms that avoid the explicit inversion of the covariance matrix. A Nystr{\"o}m-based low-rank unitary MVDR scheme \cite{rl_Nystrome} approximates the covariance matrix using a low-rank approximation and reduces the computational complexity and storage overhead.  Another approach \cite{rl_qrd} uses QR decomposition to track speech and noise variations dynamically. In \cite{rl_smi_mvdr}, an SMI-MVDR beamformer is proposed that uses recursive implementation for large arrays using Cholesky factorization and Householder transformation.

%  Distributed or parallel approaches
Distributed MVDR beamforming approaches have been developed to handle large-scale systems more effectively. A message passing algorithm \cite{rl_message_passing} is proposed to enable distributed beamforming by iteratively updating weights through local node communication. Similarly, parallel algorithms \cite{rl_parallel} and ADMM-based methods \cite{rl_distributed_mvdr} distribute the computational load across multiple processors, improving robustness and scalability while reducing latency.

%  Deep learning-based approaches
Deep learning (DL) has made significant contributions to beamforming, particularly in massive Multiple-Input Multiple Output (MIMO) systems \cite{rl_dl_survey}. Notable examples include deep adversarial reinforcement learning to enhance the capacity and performance of massive MIMO beamforming \cite{rl_dlrl1}, as well as to predict beamforming direction in mmWave Wireless Body Area Networks \cite{rl_dlrl2}. Additionally, in \cite{rl_cnn1}, Convolutional Neural Networks (CNNs) have been employed to reduce training complexity by combining supervised and unsupervised learning and for calibration state diagnosis of massive antenna arrays \cite{rl_dlrl2}.
%\vspace*{-4pt}
\section{Problem Definition}

Consider an array consisting of $M$ isotrophic antennas. Assume the array receives a signal from a target at an angle $\theta_0$ with respect to the array boresight and a collection of $L$ interfering signals from the directions $\theta_1, \theta_2, \ldots, \theta_L$ over an observation period of $m$ snapshots (number of temporal samples). The received signal vector $\mathbf{x_t} \in \mathbb{C}^M$ at the time step $t \in \{1, 2, ..., m\}$ can be expressed as,
\begin{equation}
    \label{eq1}
    \mathbf{x_t} = \mathbf{a(\theta_0)}s_0(t) + \sum_{l=1}^{L} \mathbf{a(\theta_l)}s_l(t) + \mathbf{s_n(t)}
\end{equation}
where $s_0(t) \in \mathbb{C}$ is the signal of interest, $s_l(t) \in \mathbb{C}$ is the interfering signal from the direction $\theta_l$, $s_n(t) \in \mathbb{C}^{M}$ is zero mean and spatially white Gaussian noise with variance $\sigma^2$ , and $a(\theta_i) \in \mathbb{C}^{M}$ for $i \in \{0, 1, 2, \ldots, L\}$ is the steering vector. Then the output of the beamformer can be expressed as,
\begin{equation}
    \label{eq4}
    \mathbf{y_t} = \mathbf{w}^H \mathbf{x_t},
\end{equation}
where $\mathbf{w} \in \mathbb{C}^{M}$ is the beamforming weights. According to the above model, the classic MVDR beamformer problem can be formulated as,
\begin{equation}
    \label{eq5}
    \min_{\mathbf{w}} \frac{1}{2} \mathbf{w}^H R \mathbf{w} \quad
    s.t. \quad \mathbf{w}^H \mathbf{a(\theta_0)} = 1,
\end{equation}
where $R \in \mathbb{C}^{M \times M}$ is the covariance matrix of the received signal. The solution to the above problem is given by,
\begin{equation}
    \label{eq6}
    \mathbf{w} = \frac{R^{-1} \mathbf{a(\theta_0)}}{\mathbf{a(\theta_0)}^H R^{-1} \mathbf{a(\theta_0)}}.
\end{equation}
 Practically, the covariance matrix is not known and needs to be estimated using the received signal. The sample covariance matrix at time step $n$ $(n>m)$ can be expressed as,
\begin{equation}
    \label{eq7}
    R_n = \frac{1}{m} \sum_{i=n-m +1}^{n} \mathbf{x_i}\mathbf{x_i}^H,
\end{equation}
which is the average covariance over the window of the $m$ snapshots. Then the Equation \ref{eq6} can be modified as,
\begin{equation}
    \label{eq8}
    \mathbf{w} = \frac{R_n^{-1} \mathbf{a(\theta_0)}}{\mathbf{a(\theta_0)}^H R_n^{-1} \mathbf{a(\theta_0)}}.
\end{equation}
%\vspace*{-4pt}
\section{Proposed Method}
Once we receive the signal at the time step $n+1$, recalculating the inverse of the sample covariance matrix is computationally expensive. Instead, consider the following recursive update rule for the sample covariance matrix,
\begin{equation}
    \label{eq9}
    R_{n+1} = \alpha R_n + (1-\alpha)\mathbf{x_n}\mathbf{x_n}^H,
\end{equation}
where $0 < \alpha < 1$ is the forgetting factor.

Assuming $R_n^{-1}$ is calculated using the first $n$ samples, the inverse of the covariance matrix can be updated using the Sherman-Morrison formula as,
\begin{equation}
    \label{eq10}
    {R}^{-1}_{n+1} = \frac{R_n^{-1}}{\alpha} - \frac{(1- \alpha)}{\alpha}\frac{R_n^{-1} {\mathbf{x_n}} {\mathbf{x_n}}^H R_n^{-1}}{\alpha + (1-\alpha) {\mathbf{x_n}}^H R_n^{-1} {\mathbf{x_n}}}
\end{equation}
which reduces the computational complexity to $O(M^2)$ per time step.
In order to further reduce the computational complexity, we need to have a method that does not explicitly form a covariance matrix as explicit formulation of an $M \times M$ matrix leads to $O(M^2)$ complexity. Instead, we propose the following steps that do not require the explicit formation of the covariance matrix except for the initialization step.

First, we calculate the Singular Value Decomposition (SVD) of $R$. As $R$ is Hermitian, the SVD can be expressed as,
\begin{equation}
    \label{eq11}
    R = UDU^H,
\end{equation}
where $U \in \mathbb{C}^{M \times M}$ is a unitary matrix containing the eigenvectors of $R$, $D \in \mathbb{C}^{M \times M}$ is a diagonal matrix containing the eigenvalues of $R$ in decreasing order.
We can form a low-rank approximation of $R$ by selecting the first $K$ eigenvalues and the corresponding vectors. $K$ should be selected such that $K >= L+1$. For large arrays $K << M$. The low-rank approximation can be expressed as,
\begin{equation}
    \label{eq12}
    R \approx U_KD_KU_K^H,
\end{equation}
where $U_K \in \mathbb{C}^{M \times K}$ is the first $K$ columns of $U$ and $D_K \in \mathbb{C}^{K \times K}$ is the first $K$ eigenvalues of $D$. Then we can modify the Equation \ref{eq10} as, 
\begin{equation}
    \label{eq13}
    {R}^{-1}_{n+1} \approx U_KD^{-1}_{K,n+1}U_K^H
\end{equation}
where $D^{-1}_{K,n+1}$ can be updated using the following recursive update rule,
\begin{equation}
    \label{eq14}
   \!\!\!\!\!\!\!\!\!\! D^{-1}_{K,n+1}\!\! =\!\! \frac{D^{-1}_{K,n}}{\alpha} \!- \!\frac{(1- \alpha)}{\alpha}\frac{D^{-1}_{K,n}U_K^H {\mathbf{x_n}} {\mathbf{x_n}}^H U_KD^{-1}_{K,n}}{\alpha + (1-\alpha) {\mathbf{x_n}}^HU_KD^{-1}_{K,n}U_K^H {\mathbf{x_n}}}
\end{equation}
The above update rule reduces the computational complexity to $O(MK^2)$ per time step. For large arrays as  $K << M$ the computational complexity can be approximated as $O(MK^2)$ per time step.

Finally, the beamforming weights at time step $n+1$ can be computed by substituting the updated inverse covariance matrix $ R_{n+1}^{-1}$ into Equation \ref{eq8} as,

\begin{equation}
    \label{eq16}
    \mathbf{w} = \frac{U_KD^{-1}_{K,n+1}U_K^H \mathbf{a(\theta_0)}}{\mathbf{a(\theta_0)}^H U_KD^{-1}_{K,n+1}U_K^H \mathbf{a(\theta_0)}}.
\end{equation}

The proposed method is summarized in Algorithm \ref{alg1}.

\subsection{SVD Initialization}
For MVDR beamforming, the formation of the covariance matrix requires $O(M^2)$ operations per time step and its inversion requires $O(M^3)$ operations which accounts for a total of $O(M^2 + M^3)$ operations per time step.

Note that in the proposed method, the low-rank SVD approximation of the covariance matrix is computed once during initialization, with a computational complexity of $O(M^3)$. For subsequent time steps, the MVDR beamformer is efficiently computed with a complexity of $O(MK^2)$ using the proposed algorithm.

\begin{algorithm}[t]
    \caption{Proposed Algorithm}  
        \begin{algorithmic}[1]
        \label{alg1}
        \REQUIRE Initial covariance matrix ${R}_n: M \times M$, data vectors $\mathbf{x_{n}}, \mathbf{x_{n+1}}, \mathbf{x_{n+2}}, \ldots, \mathbf{x_N}: M \times 1$, Forgetting factor $\alpha$, Low rank dimension: $K$ .

        \STATE ${R} \gets {R}_n$
  
        \STATE \text{\parbox[t]{\dimexpr\linewidth-2em}{Compute the K-rank approximation of the covariance matrix}}
        \begin{align*}
            U_KD_{K}U_K^H \gets R
      \end{align*}
  
        \FOR{$i = n$ to $N$}
            \STATE $\mathbf{x} \gets \mathbf{x_i}$
  
            \STATE \text{Update of the inverse of the submatrix ${D}$ }
            \[
            D^{-1}_K \gets \frac{D^{-1}_K}{\alpha} 
            - \frac{(1- \alpha)}{\alpha}
            \frac{D^{-1}_K U_K^H \mathbf{x} \mathbf{x}^H U_KD^{-1}_K}{\alpha + (1-\alpha) \mathbf{x}^H U_KD^{-1}_KU_K^H \mathbf{x}}
            \]
                      
            \STATE \text{Compute the Beamforming Weights}
            \[
            \mathbf{w} = \frac{U_KD_K^{-1}U_K^H \mathbf{a}(\theta)}{\mathbf{a}(\theta)^H U_KD_K^{-1}U_K^H \mathbf{a}(\theta)}
            \]

        \ENDFOR
        \end{algorithmic}
        \end{algorithm}
%\vspace*{-4pt}
\section{Experimental Results}

\subsection{Simulation Setup}

For simulations, we consider a Uniform Linear Array (ULA) with $M$ antennas where $M$ will be changed according to the experiment. The antennas are spaced half a wavelength apart. Therefore the steering vector can be expressed as  $ \mathbf{a(\theta)} = \begin{bmatrix} 1 , e^{-j\pi \sin(\theta)}, e^{-j2\pi \sin(\theta)}, \ldots, e^{-j(M-1)\pi \sin(\theta)} \end{bmatrix}^T$.

% \begin{equation}
%      \label{eq2}
%     \mathbf{a(\theta)} = \begin{bmatrix}
%     1 , e^{-j\pi \sin(\theta)}, e^{-j2\pi \sin(\theta)}, \ldots, e^{-j(M-1)\pi \sin(\theta)}

%     \end{bmatrix}^T,
% \end{equation}

\begin{figure*}[t]
    \centering
    {\includegraphics[width=\textwidth]{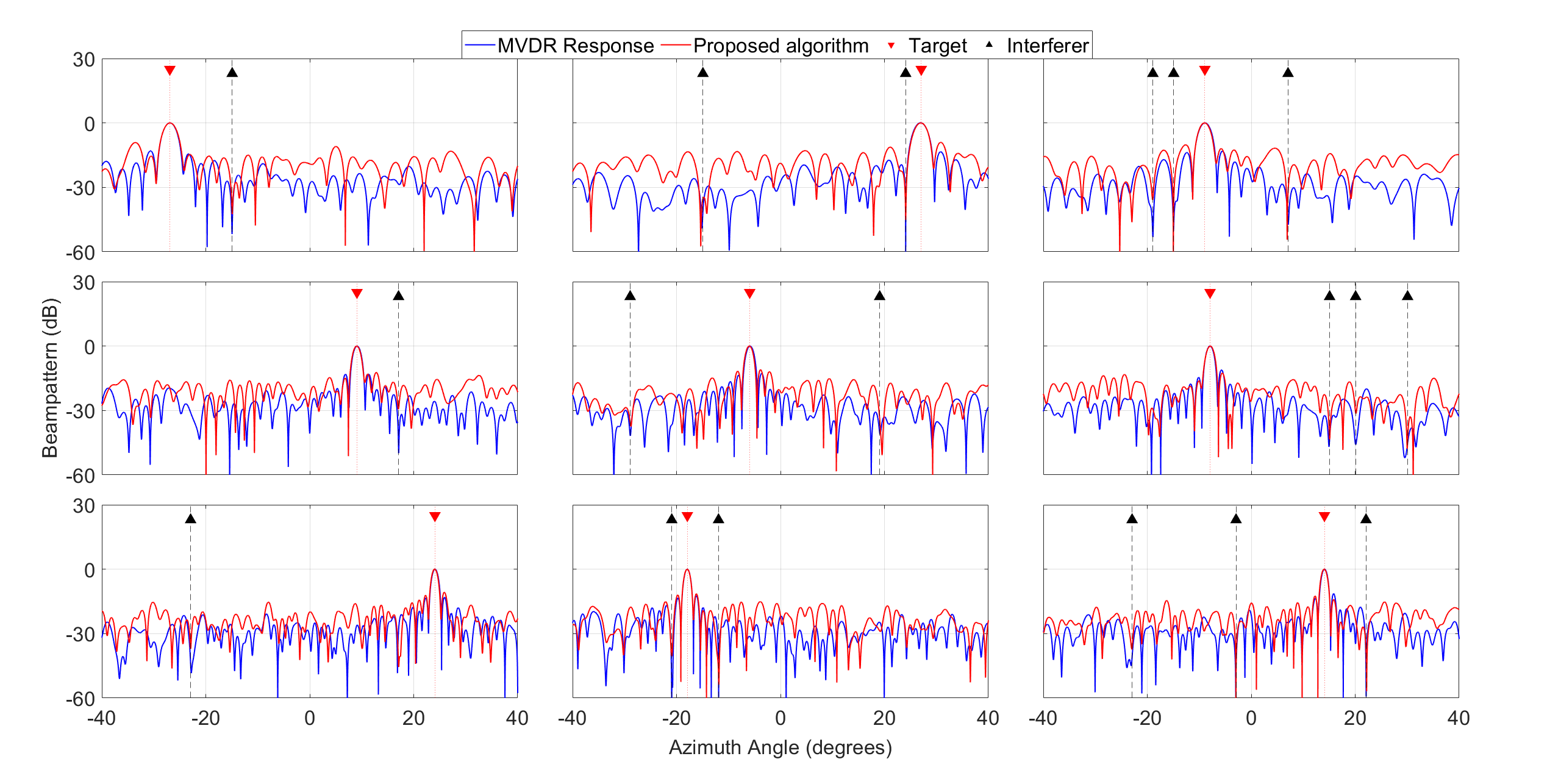}}
    \caption{{\bf Comparison of Beam patterns.} The number of antennas changes from 50, 75, and 100 from top to bottom and the number of interferers changes from 1, 2, and 3 from left to right. The proposed method closely matches the beam pattern of the conventional MVDR beamformer in placing the main lobe in the direction of the target and nulls in the direction of the interfering signals.
    }
    \label{fig:beam_patterns}
\end{figure*}

For the signal of interest, we consider an echo signal from a target located 10 km away from the array and moving tangentially to the array at a velocity of 500 m/s. The transmitted signal is a linear FM chirp with a bandwidth of 300 MHz and a pulse duration of 100 ms. The interfering signal is generated from a similar target located 10 km from the array and moving at a tangential velocity of 500 m/s. The noise is generated from a zero mean Gaussian distribution with a variance of 1. The SINR ratio is set to -10 dB.  Sampling is performed at 1 MHz. At this speed, the target moves roughly one-hundredth of a degree in about 1 ms, during which approximately 3500 samples are collected. To maintain manageable processing, we select a 1 ms observation window, during which about 1,000 samples are collected, corresponding to a smaller angular displacement of roughly 0.003$^\circ$. Therefore, we set the snapshot number to 1000. The forgetting factor $\alpha$ is set to 0.99 and the low-rank dimension $K$ is set to 10.

We initialize the DoA of the target and the interfering signals to a random value between -30 and 30 degrees. The DoA of the target and the interfering signal are updated by 0.01 degrees per 1000 samples. The beamforming weights are updated at each time step. The performance of the proposed method is evaluated in terms of computational complexity and beamforming accuracy. The computational complexity is measured by the number of operations required per time step.

\begin{table}[!t]
    \centering
    \caption{comparison of beamforming metrics between the proposed algorithm and mvdr. The proposed algorithm performs similarly to the mvdr beamformer in mlw and achieves comparable slls and null depths}
    \label{tab:beam_accuracy}
    \begin{tabular}{@{}cccccccc@{}}
    \toprule
    M   & L & \multicolumn{2}{c}{MLW  (\textdegree) $\downarrow$}  & \multicolumn{2}{c}{Null depth (dB) $\downarrow$} & \multicolumn{2}{c}{SLL (dB) $\downarrow$}  \\ 
      &  &  MVDR          & Prop         & MVDR         & Prop        & MVDR            &Prop          \\
    \midrule
    50  & 1 & 2.27          & 2.32         & -49.56         & -42.44        & -13.02            & -9.14             \\
    50  & 2 & 2.38          & 2.31         & -51.31         & -40.10         & -13.48            & -9.06             \\
    50  & 3 & 2.09          & 2.03         & -49.55         & -35.29        & -13.19            & -10.20             \\
    75  & 1 & 1.37          & 1.37         & -47.54         & -29.32        & -13.12            & -12.32            \\
    75  & 2 & 1.36          & 1.33         & -41.02         & -34.84        & -12.75            & -11.05            \\
    75  & 3 & 1.34          & 1.41         & -44.70          & -39.53        & -11.11            & -12.55            \\
    100 & 1 & 1.10           & 1.08         & -44.01         & -37.13        & -13.05            & -10.76            \\
    100 & 2 & 1.07          & 1.07         & -55.54         & -44.56        & -12.27            & -11.47            \\
    100 & 3 & 1.06          & 1.08         & -45.83         & -41.78        & -11.37            & -12.47            \\ \bottomrule
    \end{tabular}
    \end{table}

\subsection{Beamforming Accuracy}

The beamforming accuracy is evaluated in terms of the beampattern of the array. The beampattern represents the spatial response of the array to a signal arriving from different directions, with a fixed set of weights. To compute the beampattern, the array is steered across various directions, and the output power is measured for each direction. The beam pattern of the array is compared with the beampattern of the conventional MVDR beamformer. The beampatterns of the proposed method and the conventional MVDR beamformer are shown in Figure \ref{fig:beam_patterns} for $M=50, 75, 100$ and number of interferers $L=1, 2, 3$. It can be observed that the beampattern of the proposed method closely matches the beampattern of the conventional MVDR beamformer. The proposed algorithm correctly places the main lobe of the beampattern at the direction of the target and the nulls at the direction of the interfering signal.

As for quantifying the beamforming accuracy, we use the main lobe width (MLW), side lobe level (SLL), and null depth. The MLW is the angular width of the main lobe of the beam pattern. The SLLs are the power levels of the side lobes relative to the main lobe. The null depth is the power level of the nulls relative to the main lobe. The main lobe width, side lobe levels, and null depth of the proposed method and the conventional MVDR beamformer for the configurations shown in Figure \ref{fig:beam_patterns} are shown in Table \ref{tab:beam_accuracy}. The proposed algorithm performs comparably to the MVDR beamformer in terms of main lobe width (MLW), with only minor variations across configurations. While the MVDR beamformer achieves deeper nulls, the proposed method consistently achieves SLLs comparable to those of the MVDR beamformer.

\subsection{Computational Complexity}

\begin{figure}[t]
    {\includegraphics[width=.5\textwidth]{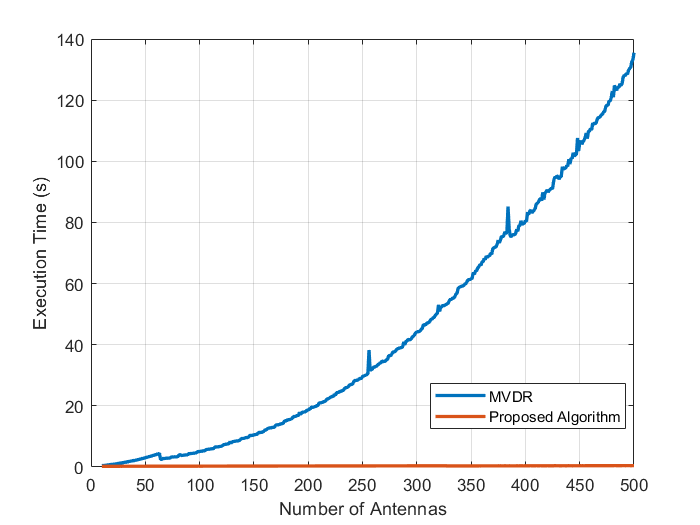}}
    \caption{{\bf Execution times. } 
    Comparison of the execution time for $M=20, 11, 12, \ldots, 500$ and $K=10$. The execution time of the proposed method scales linearly with the number of antennas, while the execution time of the conventional MVDR beamformer scales cubically with the number of antennas. This demonstrates the computational efficiency of the proposed method for large arrays.
    }
   
    \label{fig_comp_complexity}

\end{figure}

The computational complexity of the proposed method is compared with that of the conventional MVDR beamformer. The computational complexity of the conventional MVDR beamformer is $O(M^3)$ per time step. The computational complexity of the proposed method is $O(MK^2)$ per time step. For this experiment, we measured the execution time of the proposed method and the conventional MVDR beamformer for 10000-time steps for $M=10, 11, 12, \ldots, 500$ and $K=10$. The experiment was conducted on AMD EPYC 7443 24-Core Processor \cite{amd_url}. The execution time of the proposed method and the conventional MVDR beamformer is shown in Figure \ref{fig_comp_complexity}. It can be observed that the execution time of the proposed method is significantly lower than the conventional MVDR beamformer for large arrays. The execution time of the proposed method scales linearly with the number of antennas, while the execution time of the conventional MVDR beamformer scales cubically with the number of antennas.

%\vspace*{-4pt}
\section{Discussion and Limitations}

\begin{figure}[t]
    {\includegraphics[width=.5\textwidth]{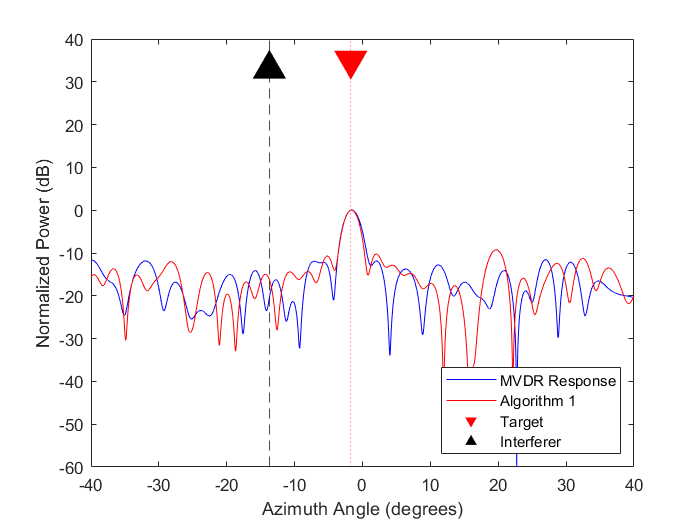}}
    \caption{{\bf Failure Mode.} Beam patterns of the proposed method and the conventional MVDR beamformer for $M=50$ and $K=10$ when the desired signal is above the noise floor: the proposed method does not place the nulls at the direction of the interfering signal as effectively as the conventional MVDR beamformer.} 
   
    \label{fig_high_snr}

\end{figure}

% Discuss the limitations of the proposed method
The proposed method is specifically designed for scenarios where the desired signal is below the noise floor. In scenarios where the desired signal is above the noise floor, the proposed method may not perform as well as the conventional MVDR beamformer. An example of this is shown in Figure \ref{fig_high_snr} where the desired signal is above the noise floor with an SNR of 10 dB. The proposed method does not place the nulls in the direction of the interfering signal as effectively as the conventional MVDR beamformer. 

\begin{figure}[t]
    {\includegraphics[width=.5\textwidth]{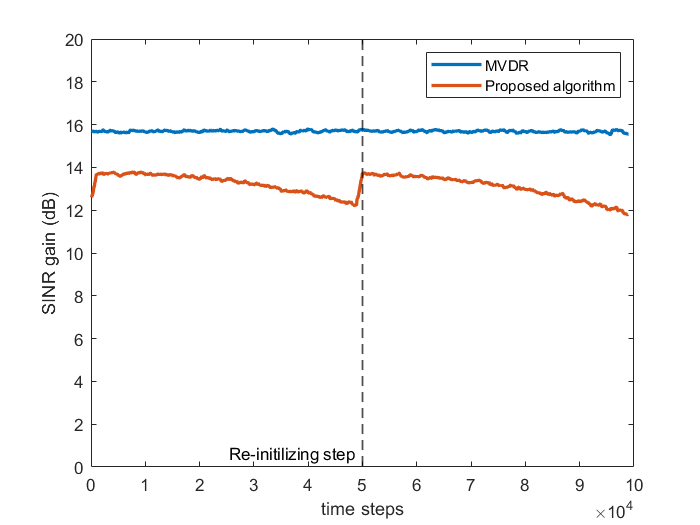}}
    \caption{{\bf SINR gain of the proposed method over the conventional MVDR beamformer for $M=100$ and $K=10$ for 100000 time steps.} The SINR gain of the proposed method decreases over time. However, the SINR gain can be restored by reinitializing the proposed method at regular intervals.}
   
    \label{fig_sinr_gain}

\end{figure}

Another limitation of the proposed method is the reduction of SINR gain when the proposed algorithm is used over long periods. SINR gain is the ratio of the output SINR to the input SINR. For MVDR and the proposed method, the output SINR can be calculated as,

\begin{equation}
    \label{eq15}
    SINR_{out} = \frac{\sigma^2_{s_0}|\mathbf{w}^H \mathbf{a(\theta_0)}|^2}{\mathbf{w}^H R \mathbf{w}}
\end{equation}
where $\sigma^2_{s_0}$ is the power of the desired signal. 
To illustrate this limitation, we calculated the average SINR gain over 10 realizations of the proposed method over the conventional MVDR beamformer for $M=100$ and $K=10$ over 100,000 time steps. The results are shown in Figure \ref{fig_sinr_gain}. It can be observed that the SINR gain of the proposed method decreases over time. However, a possible solution to this limitation is to reinitialize the proposed method at regular intervals to maintain the SINR gain. The proposed method was reinitialized for the above experiment on the 50000th time step. It can observed that the SINR gain is restored to the initial value after reinitialization.
Even with reinitialization, the proposed method offers reduced computational complexity compared to the conventional MVDR beamformer. The reinitialization step is quadratic in complexity with respect to the number of antennas, compared to the cubic complexity of the conventional MVDR beamformer.

% Efficient initialization
To further reduce the computational complexity during initialization, instead of computing the full SVD of $R$ and then truncating it to form the $K$-rank approximation, we can directly compute a low-rank approximation of $R$. This approach reduces the computational complexity from $O(M^3)$ to $O(KM^2)$, where $K$ is the rank of the approximation and $K \ll M$.
There are several methods to compute a low-rank approximation of $R$ without explicitly forming the covariance matrix. One such method is the randomized SVD \cite{halko2011finding}, which uses random projections to identify the subspace that captures the dominant singular values and vectors of $R$. By utilizing such methods, we can further reduce the computational complexity of the initialization step while maintaining the accuracy of the low-rank approximation.

% \usepackage{booktabs}

%\vspace*{-4pt}
\section{Conclusion}

In this paper, we proposed a scalable approach to MVDR beamforming for large arrays when the desired signal is below the noise floor. By incorporating the Sherman-Morrison formula and leveraging low-rank SVD approximations, we significantly reduced the complexity of updating the covariance matrix. The proposed method offers $O(MK^2)$ complexity, making it feasible for real-time applications, even with arrays consisting of hundreds or thousands of antennas.

Simulation results demonstrate that the proposed approach maintains high beamforming accuracy, effectively suppressing interference and placing nulls at the desired locations. Furthermore, the computational complexity of our method scales linearly with the number of antennas, in contrast to the cubic scaling of conventional MVDR methods. The results indicate that this approach is particularly useful in dynamic environments requiring frequent updates to the covariance matrix, as it avoids the computationally expensive direct inversion.

Future work will focus on extending the proposed method to work when the desired signal is above the noise floor. %(The current approach is prone to steer nulls at {\em any} signal above the noise floor.) 
Additionally, we plan to investigate more efficient initialization methods for the proposed algorithm to further reduce the computational complexity and improve real-time performance. Nonetheless, the proposed method offers a promising solution for real-time MVDR beamforming with large arrays.

\balance
%\vspace*{-4pt}

\bibliographystyle{IEEEtran}
\bibliography{ref}

\begin{thebibliography}{10}
\providecommand{\url}[1]{#1}
\csname url@samestyle\endcsname
\providecommand{\newblock}{\relax}
\providecommand{\bibinfo}[2]{#2}
\providecommand{\BIBentrySTDinterwordspacing}{\spaceskip=0pt\relax}
\providecommand{\BIBentryALTinterwordstretchfactor}{4}
\providecommand{\BIBentryALTinterwordspacing}{\spaceskip=\fontdimen2\font plus
\BIBentryALTinterwordstretchfactor\fontdimen3\font minus
  \fontdimen4\font\relax}
\providecommand{\BIBforeignlanguage}[2]{{%
\expandafter\ifx\csname l@#1\endcsname\relax
\typeout{** WARNING: IEEEtran.bst: No hyphenation pattern has been}%
\typeout{** loaded for the language `#1'. Using the pattern for}%
\typeout{** the default language instead.}%
\else
\language=\csname l@#1\endcsname
\fi
#2}}
\providecommand{\BIBdecl}{\relax}
\BIBdecl

\bibitem{i_largescale}
A.~M. Elbir, K.~V. Mishra, S.~A. Vorobyov, and R.~W. Heath, ``Twenty-five years
  of advances in beamforming: From convex and nonconvex optimization to
  learning techniques,'' \emph{IEEE Signal Processing Magazine}, vol.~40,
  no.~4, pp. 118--131, 2023.

\bibitem{i_large_issues}
Y.~Xu, E.~G. Larsson, E.~A. Jorswieck, X.~Li, S.~Jin, and T.-H. Chang,
  ``Distributed signal processing for extremely large-scale antenna array
  systems: State-of-the-art and future directions,'' \emph{arXiv preprint
  arXiv:2407.16121}, 2024.

\bibitem{i_beamforming2}
\BIBentryALTinterwordspacing
B.~D. Van and K.~M. Buckley, ``{Beamforming: a versatile approach to spatial
  filtering},'' \emph{IEEE ASSP Mag.}, vol.~5, pp. 4--24, Apr. 1988. [Online].
  Available: \url{https://ieeexplore.ieee.org/abstract/document/665/}
\BIBentrySTDinterwordspacing

\bibitem{i_beamforming}
W.~Liu and S.~Weiss, \emph{Wideband beamforming: concepts and
  techniques}.\hskip 1em plus 0.5em minus 0.4em\relax John Wiley \& Sons, 2010.

\bibitem{i_beamforming4}
\BIBentryALTinterwordspacing
N.~A. Gumerov, B.~Zhi, and R.~Duraiswami, ``{Sequential Direction Detection for
  Sound Scene Analysis},'' in \emph{{2018 IEEE International Conference on
  Acoustics, Speech and Signal Processing (ICASSP)}}, Apr. 2018, pp.
  6807--6811. [Online]. Available:
  \url{http://dx.doi.org/10.1109/ICASSP.2018.8462314}
\BIBentrySTDinterwordspacing

\bibitem{i_beamforming3}
\BIBentryALTinterwordspacing
A.~V.~D. Veen, ``{Algebraic methods for deterministic blind beamforming},''
  \emph{Proc. IEEE Inst. Electr. Electron. Eng.}, vol.~86, pp. 1987--2008,
  1998. [Online]. Available:
  \url{https://ieeexplore.ieee.org/abstract/document/720249/}
\BIBentrySTDinterwordspacing

\bibitem{i_mvdr}
H.~L. Van~Trees, \emph{Optimum array processing: Part IV of detection,
  estimation, and modulation theory}.\hskip 1em plus 0.5em minus 0.4em\relax
  John Wiley \& Sons, 2002.

\bibitem{rl_smi_mvdr}
V.~V. Zaharov and M.~Teixeira, ``{SMI-MVDR} beamformer implementations for
  large antenna array and small sample size,'' \emph{IEEE Transactions on
  Circuits and Systems I: Regular Papers}, vol.~55, no.~10, pp. 3317--3327,
  2008.

\bibitem{i_sherman}
J.~Sherman and W.~J. Morrison, ``Adjustment of an inverse matrix corresponding
  to a change in one element of a given matrix,'' \emph{The Annals of
  Mathematical Statistics}, vol.~21, no.~1, pp. 124--127, 1950.

\bibitem{i_svd}
G.~Strang, \emph{Linear Algebra and Its Applications 4th ed.}, 2012.

\bibitem{rl_Nystrome}
S.~Jiang, M.~Jin, S.~Liu, and Z.~Lin, ``A {Nystr{\"o}m-based} low-rank unitary
  {MVDR} beamforming scheme,'' \emph{Signal Processing}, vol. 220, p. 109433,
  2024.

\bibitem{rl_qrd}
A.~Barnov, V.~B. Bracha, and S.~Markovich-Golan, ``{QRD} based {MVDR}
  beamforming for fast tracking of speech and noise dynamics,'' in \emph{2017
  IEEE Workshop on Applications of Signal Processing to Audio and Acoustics
  (WASPAA)}.\hskip 1em plus 0.5em minus 0.4em\relax IEEE, 2017, pp. 369--373.

\bibitem{rl_message_passing}
R.~Heusdens, G.~Zhang, R.~C. Hendriks, Y.~Zeng, and W.~B. Kleijn, ``Distributed
  {MVDR} beamforming for (wireless) microphone networks using message
  passing,'' in \emph{IWAENC 2012; International Workshop on Acoustic Signal
  Enhancement}.\hskip 1em plus 0.5em minus 0.4em\relax VDE, 2012, pp. 1--4.

\bibitem{rl_parallel}
P.~Sinha, A.~D. George, and K.~Kim, ``Parallel algorithms for robust broadband
  {MVDR} beamforming,'' \emph{Journal of Computational Acoustics}, vol.~10,
  no.~01, pp. 69--96, 2002.

\bibitem{rl_distributed_mvdr}
W.~Zhang, J.~Lin, X.~Wu, and Y.~Pan, ``A distributed approach to robust minimum
  variance distortionless response beamforming in large-scale arrays,''
  \emph{IET Communications}, vol.~17, no.~8, pp. 950--959, 2023.

\bibitem{rl_dl_survey}
H.~Al~Kassir, Z.~D. Zaharis, P.~I. Lazaridis, N.~V. Kantartzis, T.~V.
  Yioultsis, and T.~D. Xenos, ``A review of the state of the art and future
  challenges of deep learning-based beamforming,'' \emph{IEEE Access}, vol.~10,
  pp. 80\,869--80\,882, 2022.

\bibitem{rl_dlrl1}
T.~Maksymyuk, J.~Gazda, O.~Yaremko, and D.~Nevinskiy, ``Deep learning based
  massive {MIMO} beamforming for {5G} mobile network,'' in \emph{2018 IEEE 4th
  International Symposium on Wireless Systems within the International
  Conferences on Intelligent Data Acquisition and Advanced Computing Systems
  (IDAACS-SWS)}.\hskip 1em plus 0.5em minus 0.4em\relax IEEE, 2018, pp.
  241--244.

\bibitem{rl_dlrl2}
H.~Ngo, H.~Fang, and H.~Wang, ``Deep learning-based adaptive beamforming for
  {mmWave} wireless body area network,'' in \emph{GLOBECOM 2020-2020 IEEE
  Global Communications Conference}.\hskip 1em plus 0.5em minus 0.4em\relax
  IEEE, 2020, pp. 1--6.

\bibitem{rl_cnn1}
S.~Lu, S.~Zhao, and Q.~Shi, ``Learning-based massive beamforming,'' in
  \emph{GLOBECOM 2020-2020 IEEE Global Communications Conference}.\hskip 1em
  plus 0.5em minus 0.4em\relax IEEE, 2020, pp. 1--6.

\bibitem{amd_url}
AMD, ``{AMD EPYC} 7443,''
  \url{https://www.amd.com/en/products/processors/server/epyc/7003-series/amd-epyc-7443.html},
  accessed: 2024-10-28.

\bibitem{halko2011finding}
N.~Halko, P.-G. Martinsson, and J.~A. Tropp, ``Finding structure with
  randomness: Probabilistic algorithms for constructing approximate matrix
  decompositions,'' \emph{SIAM review}, vol.~53, no.~2, pp. 217--288, 2011.

\end{thebibliography}

\end{document}